# A Framework for Constraint-Based Deployment and Autonomic Management of Distributed Applications

Alan Dearle, Graham Kirby and Andrew McCarthy
*School of Computer Science, University of St Andrews, St Andrews, Fife KY16 9SS, Scotland*
*{al, graham, ajm}@dcs.st-and.ac.uk*

## Abstract

*We propose a framework for deployment and subsequent autonomic management of component-based distributed applications. An initial deployment goal is specified using a declarative constraint language, expressing constraints over aspects such as component-host mappings and component interconnection topology. A constraint solver is used to find a configuration that satisfies the goal, and the configuration is deployed automatically. The deployed application is instrumented to allow subsequent autonomic management. If, during execution, the manager detects that the original goal is no longer being met, the satisfy/deploy process can be repeated automatically in order to generate a revised deployment that does meet the goal.*

## 1. Introduction

In [1], Kephart & Chess describe an autonomic system as possessing the following aspects of self-management:

- self-configuration
- self-optimization
- self-healing
- self-protection

This is illustrated in Figure 1, which shows a *managed element* and its autonomic lifecycle. The element is associated with an autonomic manager that attempts to maintain some high-level objective for the element. The behaviour of the element is continually monitored and analysed. When this deviates sufficiently from the objective, the manager plans and executes a change to the element in order to restore the desired behaviour.

In this paper we describe a framework for autonomic management of deployment and configuration of distributed applications. The managed elements are collections of components making up a distributed application. We assume that the component granularity is relatively large and that components are not nested.

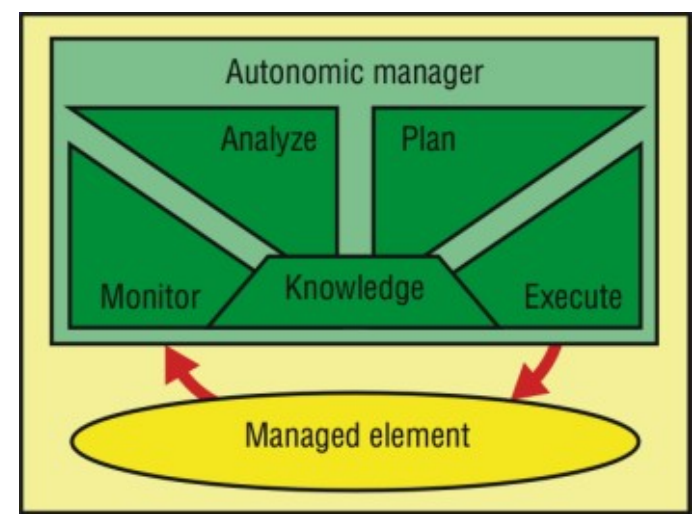

**Figure 1. A managed element (from [1])**

We identify two separate but closely related problems: the initial deployment of an application, and its subsequent evolution in the face of host failures and other perturbations. Both are too complex in large applications to be handled by a human operator. In our framework both are controlled automatically, driven by a high-level configuration goal specified by the administrator at the outset. We thus address specifically the first and third of Kephart & Chess' issues: self-configuration and self-healing. Although not addressed in this paper, we believe that self-optimization and self-protection can be also be accommodated within this framework.

It is our thesis that to implement such an autonomic deployment and configuration cycle, we require:

1. a mechanism for deploying components
2. a language to describe how the application is intended to be structured
3. an autonomic management engine capable of
   - identifying a valid configuration of the application
   - deploying the configuration into a distributed environment
   - modifying the deployed application to maintain the specified intended structure in the face of changing circumstances

Some mechanism is required for deploying and redeploying components in possibly remote locations. We advocate the use of *bundles*, which were developed by us in the project Computation in Geographically

Appropriate Locations (Cingal) [2, 3]. Bundles permit XML-encoded closures of code and data to be pushed and executed in remote locations. Cingal-enabled hosts provide a light-weight runtime and security infrastructure, written in pure Java, necessary to support the execution of bundles.

In order to describe how an application is intended to be structured, we propose a domain-specific constraint-based language. This describes configuration goals in terms of resources including software components and physical hosts, relationships between hosts and components, and constraints over these. From such a configuration goal it is possible to:

- deploy components using the available physical resources
- configure monitoring software to assess whether the executing application continues to obey the constraints specified in the description
- configure software for automatically evolving the application in response to constraint violations arising from changes in the environment

There are several levels at which a deployed application may be evolved. The simplest, on which we concentrate here, involves evolution of the configuration in order to maintain a previously specified goal. Thus the configuration evolves whilst the high-level configuration goal remains the same. We term this *autonomic evolution*, and consider it to be fundamental to the autonomic management of distributed applications. Our aim is for this style of evolution to take place automatically whenever required.

A second level of evolution is needed when the high-level goal itself changes, due to a change in application requirements. Our framework handles both levels of evolution in the same way, treating the first as a special case of the second in which the goal remains fixed. In both cases an ongoing autonomic cycle, as shown in Figure 1, repeatedly attempts to solve the current constraint problem, deploys the resulting configuration, and monitors the deployment to determine when to repeat the sequence.

## 2. Related languages and systems

The Cingal system supports the deployment of distributed applications in geographically appropriate locations. It provides mechanisms to execute and install components, in the form of *bundles*, on remote machines. A bundle is the only entity that may be executed in Cingal and consists of an XML-encoded closure of code and data and a set of bindings naming the data. Cingal-enabled hosts contain appropriate security mechanisms to ensure malicious parties cannot deploy and execute harmful agents, and to ensure that deployed components do not interfere with each other either accidentally or maliciously. Cingal components may be written using standard programming languages and programming models. When a bundle is received by a Cingal-enabled host, provided that the bundle has passed a number of checks, the bundle is *fired*, that is, it is executed in a security domain (called a *machine*) within a new operating system process. Unlike processes running on traditional operating systems, bundles have a limited interface to their local environment. The repertoire of interactions with the host environment is limited to: interactions with a local store, the manipulation of bindings, the firing of other bundles, and interactions with other Cingal processes. The approach described in this paper exploits much of the technology provided by Cingal.

A number of languages have been developed to describe software architectures, including [4-6]. Typical of these is Acme [7], which is intended to fulfil three roles: to provide an architectural interchange format for design tools, to provide a foundation for the design of new tools and to support architectural modelling. The Acme language supports the description of components joined via connectors which provide a variety of communication styles. Components and connectors may be annotated with properties that specify attributes such as source files and degrees of concurrency, etc. Acme also supports a logical formalism based on relations and constraints which permits computational or run-time behaviour to be associated with the description of architectures. Acme does not however support the deployment of systems from the architectural descriptions, nor does it encompass physical computation resources.

The ArchWare ADL [8] is based on higher-order $\pi$-calculus, and is aimed at specifying active architectures, in which the architectural description of an application evolves in lock-step with the application itself. The language supports a reversible *compose* operator that allows components to be assembled from other components, and later decomposed and recomposed to permit evolution. Decomposition operates at a fine-grain, and it is possible to decompose a component into constituent parts without losing encapsulated state. This is achieved using *hyper-code*, which provides a reified form for both code and data. In comparison, the evolution in our framework operates at a coarser grain, and we assume that one component may be completely replaced by another with no common state. Further, the ArchWare ADL focuses on software architecture and does not address physical deployment.

The Active Pipes approach [9] encompasses the notions of machines and processes which transform data in an active network. The idea is to map a high level pipeline of software components onto physical network resources. As the authors state in their paper, “it is necessary to have a general scheme of specifying application requirements that is expressive enough to describe typical application scenarios while simple enough to be used effectively”. In our framework we aim to combine this approach with the notion of an ADL to encompass hardware and software components.

Constraint programming models problems by declaring a set of variables with finite domains and constraints between values of these variables. Instead of writing an imperative program to provide a result, the user invokes a search algorithm to find a solution which satisfies the constraints specified by the user. A number of constraint programming and solving systems exist. We believe that the lack of domain-specific syntax in such systems makes them unsuitable for specifying the high-level configuration goal in an autonomic application. However, they are applicable as constraint solvers when used in conjunction with a domain-specific language.

For example, ECLiPSe [10] is a constraint logic programming system with syntax similar to Prolog, supplied with a number of constraint solvers and libraries. JSolver [11] is a commercial Java library which provides constraint satisfaction functionality, while Cream [12] is a simpler open-source library. Any such systems could be employed in our framework.

In Section 4 we describe a new domain-specific constraint language, **Deladas** (DEclarative LAnguage for Describing Autonomic Systems), which is suitable for specifying autonomic systems and may be used to drive the deployment and evolution process.

## 3. Our Ap

Our gene
plication ad
terms of res
deployment.

nents and physical hosts on which these components may be installed and executed. Constraints operate over aspects such as the mapping of components to hosts and the interconnection topology between components.

We assume that the distributed application can be structured as encapsulated components, each with its own thread of control. The granularity of components is intended to be large, so that a relatively small number of components execute on each host. The components must be capable of recovering their own state if necessary, for example, in the event of a host crash. In our current prototype, components communicate with one another via asynchronous channels, but the approach could be extended in a straight-forward manner to support other styles such as RPC. We also assume that the application contains application-level protocols that cope with the disconnection and reconnection of channels to different platforms and servers. One such technology is the *half session* abstraction described by Strom and Yemeni [13].

The cycle shown in Figure 2 is controlled by the Autonomic Deployment and Management Engine (ADME). In order to produce a concrete deployment of the application, the ADME attempts to **satisfy** the goal, specified by the administrator in the Deladas language. The engine includes a Deladas parser and constraint solver. The result of the attempted goal satisfaction is a set of zero or more solutions. Each solution is in the form of a **configuration**, which describes a particular mapping of components to hosts and interconnection topology that satisfies the constraints. Configurations are encoded in XML documents known as *Deployment*

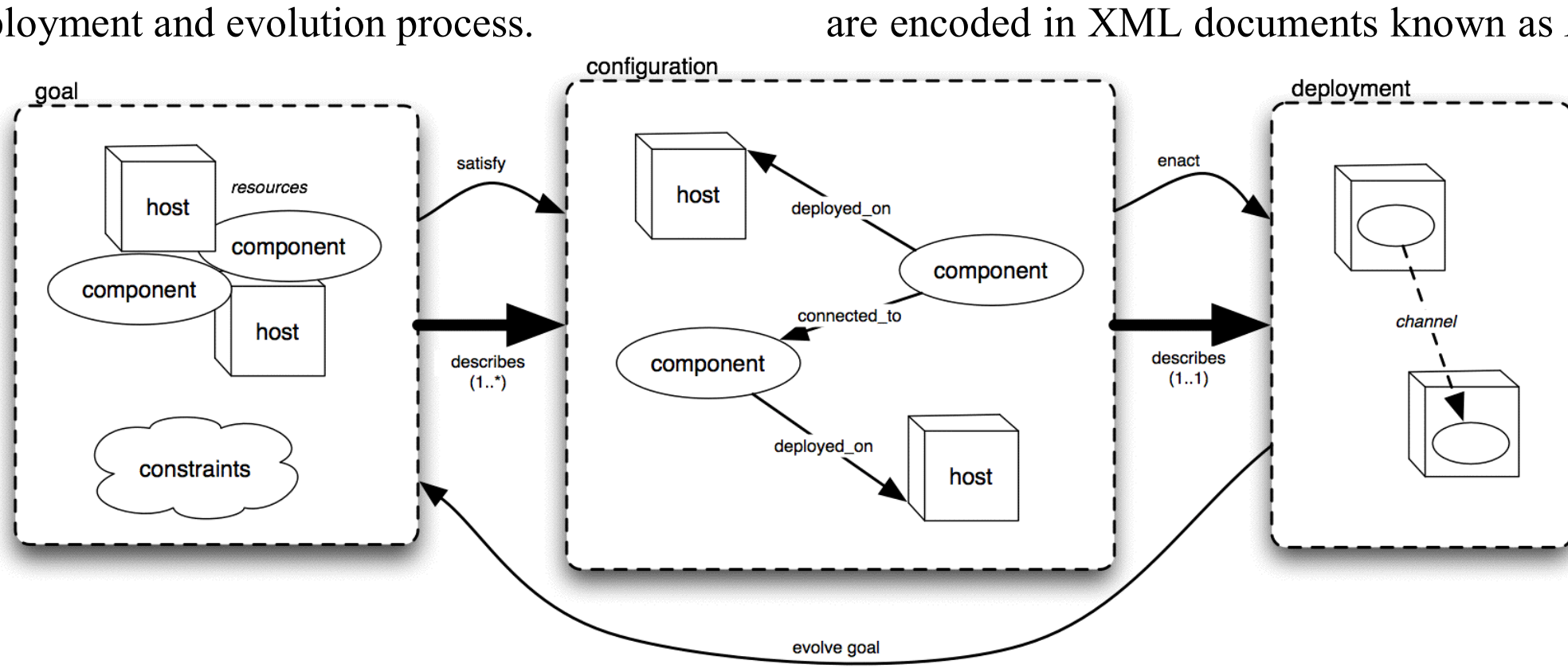

**Figure 2. Refined autonomic cycle**

If a configuration can be found[1], it is **enacted** by ADME to produce a running deployment of the appli-cation. This is facilitated using Cingal, and the Cingal infrastructure must already be installed on each of the hosts involved. From a configuration expressed as a DDD, ADME generates a collection of bundles which perform installation, instantiation and wiring of the components. Once these bundles have been fired on the appropriate hosts, the application is fully deployed in its initial configuration. This process is described in detail in [14].

[1] The ADME may be configured to use the first configuration found, or to allow the administrator to choose among multiple configurations.

The autonomic aspect of this approach is that the deployed application is instrumented with probes to monitor its execution. Events generated by the probes are sent to the ADME, which may decide that the deployment no longer satisfies the original goal, for example if a component or host fails. In this case the ADME evolves the goal to take account of changed resource availability—for example, removing failed hosts and perhaps adding new hosts that may now be available—and initiates the **satisfy**/**enact** cycle again. This attempts to find a new solution of the constraints that combines existing and new components, and to enact this in an efficient manner. Assuming that such a new configuration can be found and deployed, the system has reacted automatically and appropriately to a change in the application's environment. The cycle may continue indefinitely. This process is described in more detail in Section 5.

The nature of the probes required to monitor the application depends on the constraints specified in the goal. At the simplest level the constraints operate over just the component/host topology, and for this, simple probes are sufficient. Where more complex probes are required, this can be deduced by ADME from the specified constraints. For example, constraints can operate over the latency or bandwidth of a channel, the degree of replication of a component, or the mean availability of a host. Each of these dynamic aspects requires a specialised probe. We view Deladas as a core language that may be extended to incorporate new constraint types and associated probes.

This style of autonomic application evolution can be achieved without human intervention. The framework described above also accommodates the need for more wide-ranging evolution. For example, in addition to changes in the application's environment, changes may occur in the enterprise that the application supports; examples include changes in legal or financial regulations, or mergers of organisations. These may require manual revision of the deployment goal, including changes to the constraints.

## 4. Initial Deployment

In this section we explore, using an example, the use of Deladas to describe the resources and constraints described in the last section. The language belongs to the family of architectural description languages (ADLs). Unlike some ADLs, Deladas does not contain any computational constructs, and programs that perform computation cannot be written in it; it is purely declarative and descriptive.

We believe that Deladas' constraint style of deployment specification gives it a relative simplicity compared with more explicit styles, making it suitable for the specification of relatively large application deployments. This is especially important when the deployment is to be recomputed repeatedly in an autonomic cycle.

Deladas is used to define systems and constraints over them. The types supported are: *component*, *host* and *constraintset*. The type *component* is used to describe software components at a high level. Components, like many of the types in Acme, have associated attributes. The mandatory attributes for components are *bundles* and *ports*. Bundles are used to define the code and static data of the components. Ports are used to define communication channels between components. The type *host* is used to describe a resource on which components can be deployed. Attributes of hosts include IP-address, ownership, platform type, etc.

The type *constraintset* is a high level constraint-based specification of the invariants that pertain to a system. A *constraintset* constrains the way in which the system is realised, for example how processes are placed on machines and how the processes are wired up. *Constraintsets* are used to yield an initial configuration that might be deployed, and also to constrain deployments in the face of change. In the future we envisage extending the *constraintsets* described here to include other aspects such as bandwidth and geopolitical constraints.

To illustrate the use of *constraintsets*, we use an example drawn from the peer-to-peer domain, in which *clients* connect to *routers*. Figure 3 shows one particular configuration that satisfies the deployment goal, expressed as a Deladas *constraintset*, shown in Figure 4. In the configuration shown in Figure 3, the six hosts, labelled *h1* to *h6*, each contain a single component, labelled *C* for client and *R* for router. The components are connected via uni-directional channels, which are attached to particular ports on each component.

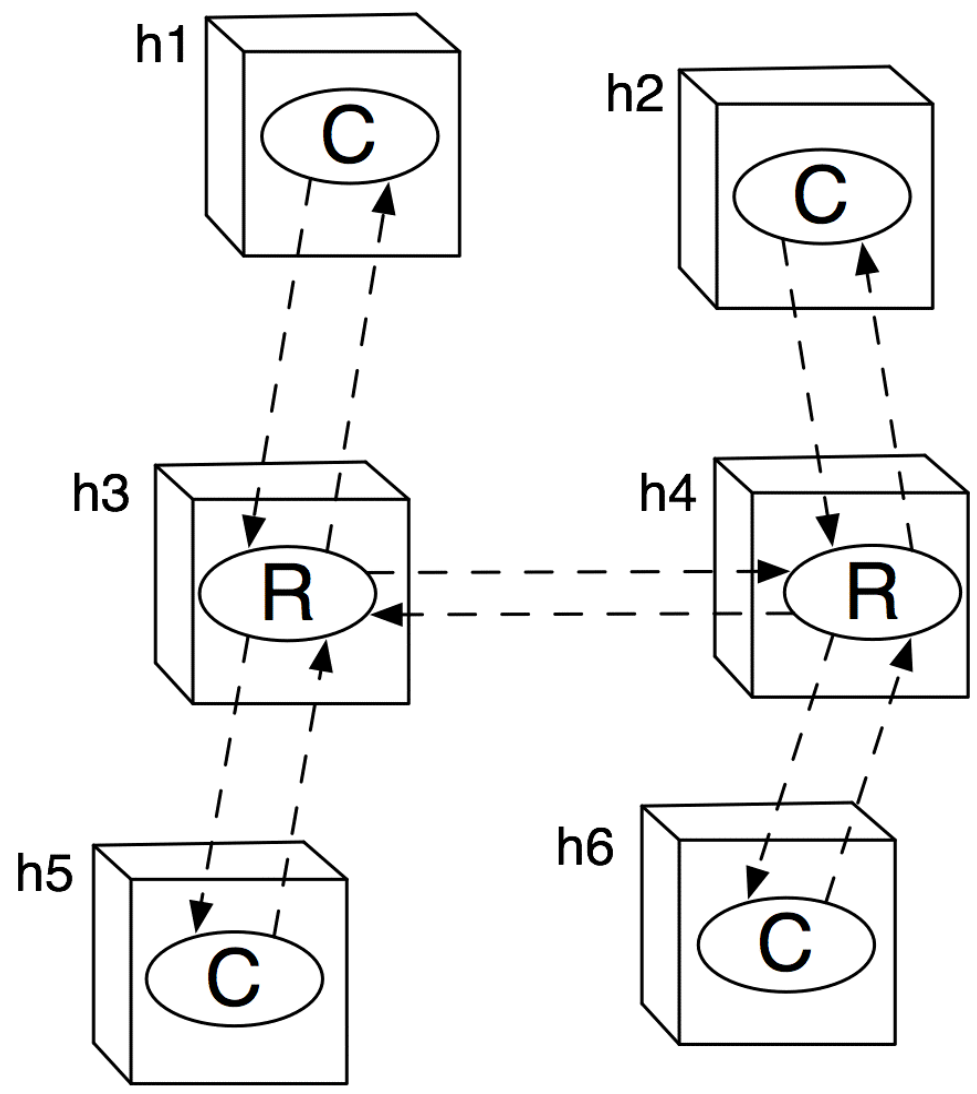

**Figure 3. Example configuration**

We now describe in more detail the Deladas *constraintset* shown in Figure 4. Given a set of resources specified in Deladas and comprising *components* and *hosts*, the *constraintset* might describe none, one or many possible configurations. It is easily possible to write Deladas *constraintsets* that are internally inconsistent and therefore specify no valid configurations, irrespective of resources. The writing of appropriate *constraintset* definitions is likely to remain difficult, and we envisage that *constraintsets* for common architectural patterns might be available off the shelf, presenting the opportunity for high level architectural reuse and specialisation.

In this example, the *constraintset* contains five constraint clauses. These clauses operate over two types of component named *Router* and *Client*. It is not necessary to specify the concrete types of these components but it is possible to infer that, in order to satisfy the constraints, the component *Router* must have ports named *rin*, *rout*, *cin* and *cout*. The constraints are written in first-order logic and specify (in sequence) that:

- hosts run an instance of a router and/or a client
- every client connects to at least one router via the *out* and *in* ports on the client and the *cin* and *cout* ports on the router
- there are at most two clients for every router
- every router is connected to at least one other router via their *rin* and *rout* ports
- routers are strongly connected

Note that if two clients are connected to a router, routers require a separate *cin* and *cout* port per client.

```
constraintset randc = constraintset {

   // 1 router or client per host
   forall host h in deployment (
      card(instanceof Router in h) = 1 or
      card(instanceof Client in h) = 1
   )

   // every client connects to at
   // least 1 router
   forall Client c in deployment (
      exists Router r in deployment (
         c.out connectsto r.cin
         c.in connectsto r.cout
      )
   )

   // every router connects to at
   // most 2 clients
   forall Router r in deployment (
      card(Client c connectedto r) <= 2
   )

   // every router connects to at
   // least 1 other router
   forall Router r1 in deployment (
      exists Router r2 in deployment (
         r1.rout connectsto r2.rin
         r1.rin connectsto r2.rout
         r1 != r2
      )
   )

   // routers are reachable from each other
   forall Router r1,r2 in deployment (
      reachable(r1, r2)
   )
}
```

**Figure 4. Example Deladas constraintset**

Figure 5 shows an example Deladas specification of resources that might be given to the solver in order to obtain a deployment. This specification defines the components *Client* and *Router*. The specification of *Client* includes the bundle containing code and static data, and defines two ports named *in* and *out*. The port definition of *Router* states that routers may have a multiplicity of connections, designated by the bracket notation. This variadicity is missing in many ADLs, preventing the specification and generation of architectures like the example architecture used in this paper.

```
component Client(
   code = "file:///D:ClientBundle.xml",
   ports = {in, out}
)
component Router(
   code = "http://deladas.org/RBundle.xml",
   ports = {cin[], cout[], rin[], rout[]}
)
host h1 = host(ipaddress = "192.168.0.1")
...
host h6 = host(ipaddress = "192.168.0.6")
```

**Figure 5. Example Deladas resources**

## 5. Autonomic Cycle

Here we describe in more detail the autonomic cycle first described in Section 3. We assume that the clients and routers described in Figures 4 and 5 have been deployed in the topology shown in Figure 3, which is compliant with the Deladas constraints. Figure 6 shows part of this deployment in more detail. Each component executes within a Cingal-supported machine as a separate operating system level process. For each host running a component, the system deploys another component called the Autonomic Management Process (AMP). This task is responsible for monitoring the health of each of the deployed components running on that host. The overall orchestration of the deployed system is the responsibility of an instance of the ADME. It is unimportant whether this is the same instance that caused the original deployment of the architecture, or not. To avoid ambiguity we will call the instance of the ADME performing the orchestration the Monitoring ADME (MADME). The MADME holds the knowledge required for the autonomic cycle in the form of the resources (components and hosts) and the constraints over those resources.

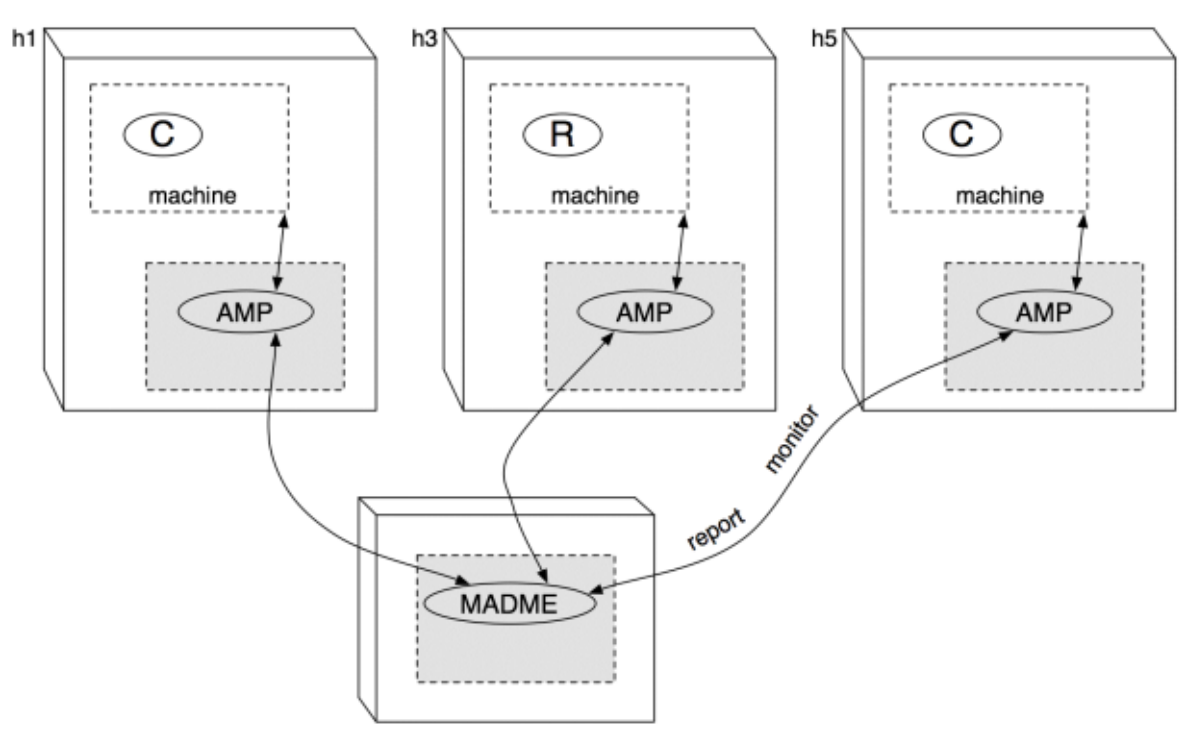

**Figure 6. Components for autonomic management**

It is now possible to see how the autonomic cycle shown in Figure 2 is implemented. An instance of the ADME solves the constraints and the resulting architecture is enacted by ADME to produce a running deployment. This deployment may include a new MADME process, or the ADME instance may become the MADME for the deployment. When events are received by the MADME that indicate invalidation of the constraints, the MADME attempts to find a new solution to the constraints. We have glossed over two details—how the changes are detected and how stability of the system is maintained.

When a system is deployed, in addition to the resources and constraints specified in Deladas, the MADME has knowledge of the identity of the Cingal machines executing the components, and of the AMP processes. Each Cingal machine running a component knows of its local AMP process, which is configured with knowledge of the MADME. To illustrate how the autonomic cycle is initiated we will consider two possible failures: the failure of the router process running on host *h3*, and the failure of the entire node *h3*.

In the event of the router process running on *h3* failing (say due to a heap overflow), various different entities can potentially observe the failure: the connected clients running on hosts *h1* and *h5*, the connected router running on host *h4*, the MADME, or the collocated AMP. The failures can be detected either by the loss of a connection to other processes or by using heartbeats between the components. The entities observing the failure are commonly known as *failure suspectors* and the approach to recovery advocated here is perhaps first due to Birman [15].

In practice, being able to determine which component has failed in the face of unreliability is notoriously difficult, and there exists a large body of work on unreliable failure suspectors, e.g. [16, 17]. For the purposes of this paper we assume that we can reliably determine which hosts and/or components have failed, and that the failures will be reported to the MADME.

If a failure has been reported by the collocated AMP, the MADME can trivially determine that it is the process hosting the router and not the host that has failed. In this case the MADME can instantiate a new router instance on node *h3* using a subset of the functionality used to initially create it. If the entire *h3* node fails, the MADME is required to find a new solution to the constraints. However, before examining how this is performed, the issue of stability of constraint solutions must be addressed.

The solution to the placement of clients and routers shown in Figure 3 is one of many possible solutions to the constraints given in the Deladas specification. Other solutions may be trivially found by hosting the routers on hosts *h1* and *h2* for example. When the MADME is required to find a new solution to the specified constraints, it is desirable to minimise the redeployment of processes between hosts. Before attempting to find a new solution to the general problem, as it did when the initial deployment was determined, the MADME therefore attempts to solve a more constrained problem. In this case, the problem is formed from the original constraints and resources, and the bindings surviving from the original deployment, comprising *R* to *h4*, *C* to *h1*, *C* to *h2*, *C* to *h5* and *C* to *h6*. If no solution can be found to this problem, the extant bindings are progressively removed from the description until a solution can be found.

Like the original attempt to find a solution, there is always the possibility that no solution may be found. If no solution can be found, a constraint error is issued by the MADME. This can be delivered via a variety of mechanisms.

In the situation where the host *h3* fails completely, the MADME might find the solution shown in Figure 7.

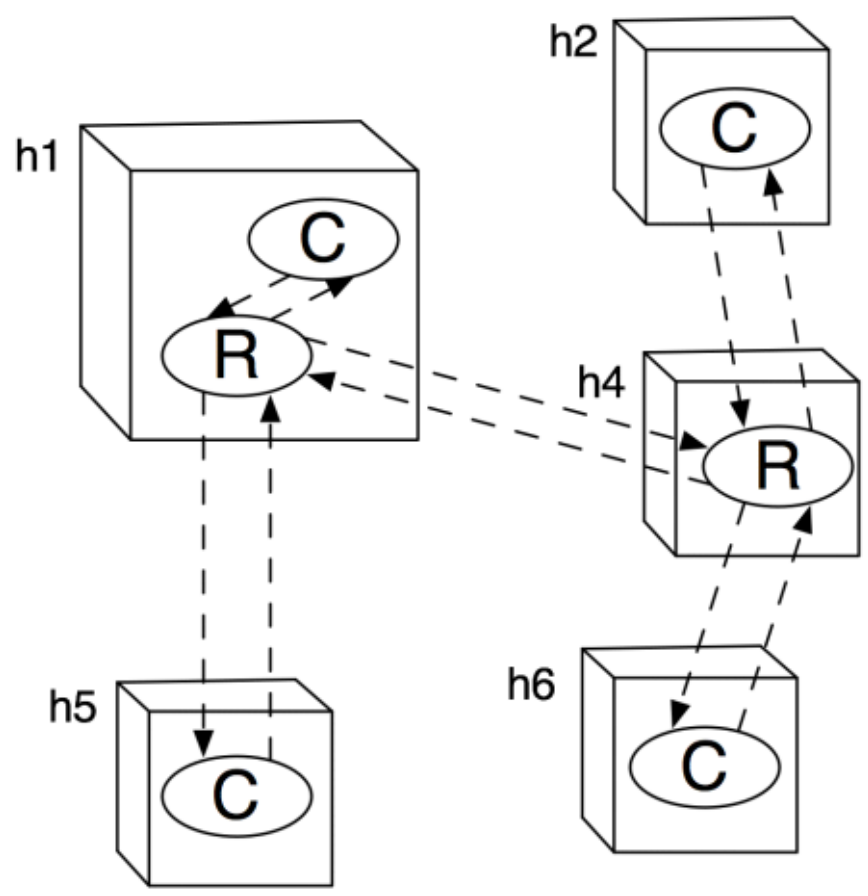

**Figure 7. Evolved configuration**

Thus far, the autonomic processes described have not included any human intervention. However, as discussed earlier, changes may occur in the enterprise that the application supports, requiring manual revision of the deployment goal, including changes to the constraints. In order to accommodate such changes, mechanisms are required whereby the resources and constraints may be changed by human agents. This may be achieved via direct interaction with the MADME.

The situations where resources are changed are similar to that where evolution is forced due to some failure. Changes initiated by a human are richer than those that are machine-initiated since resources can be added as well as removed. However, the changing of constraints cannot occur without human intervention. To accommodate these changes, the MADME presents five methods (as Web services) to the outside world, shown in Figure 8.

```
String getResources();
String getConstraints();
String getConfig();
String[] satisfy(String config,
                 String resources,
                 String constraints);
void enact(String config);
```

**Figure 8. MADME external interface**

The first three methods are selectors enabling the Deladas resources and constraints and the DDD describing the deployment to be obtained. The *satisfy* method allows new constraints, resources and existing deployed components to be specified in order to accommodate some enterprise-level change. The *config* parameter may be null, corresponding to the initial deployment problem. The *satisfy* method returns a collection of DDDs compliant with the specified constraints. The *enact* method performs enactment as described earlier. This may require extant processes to be terminated and redeployed elsewhere.

## 6. Status and further work

The main constituents of the framework described in this paper are:

- the Deladas language;
- the constraint solver;
- the component deployment mechanism;
- the monitoring infrastructure; and
- the ADME autonomic manager

Of these, the component deployment mechanism is fully implemented, based on the Cingal system [3]. It takes an XML description of a configuration and deploys it on a set of Cingal-enabled hosts. We have implemented the Deladas language, and are investigating several constraint programming tools including ECLiPSe [10], JSolver [11] and Cream [12]. The monitoring infrastructure and autonomic manager will be developed once the initial *satisfy/enact* functionality is operational. We would hope to have a full prototoype implementation completed by the time of the conference.

We plan to evaluate the basic utility of the framework initially by deploying several distributed applications such as a load-balanced web server and a publish/subscribe network onto a Beowulf cluster, and forcing various types of host and component failure. Longer term we will investigate the scalability of the framework, in particular the tractability of the constraint solving part, and experiment with extensibility in terms of the constraints and monitoring infrastructure that can be incorporated.

## 7. Conclusions

We believe that autonomic management of distributed application deployment will become essential as the scale and complexity of applications grow. This paper has outlined a framework to support the initial deployment and subsequent autonomic evolution of distributed applications in the face of perturbations such as host and link failure, temporary bandwidth problems, etc. The knowledge required for autonomic management is specified in the form of a set of available hardware and software resources and a set of constraints over their deployment. We postulate that it is feasible to implement an autonomic manager that will automatically evolve the deployed application to maintain the constraints while it is in operation. We are currently working on an implementation to enable us to test this assertion.

## Acknowledgements

This work is supported by EPSRC Grants GR/M78403 "Supporting Internet Computation in Arbitrary Geographical Locations", GR/R51872 "Reflective Application Framework for Distributed Architectures" and GR/S44501 "Secure Location-Independent Autonomic Storage Architectures", and by EC Framework V IST-2001-32360 "ArchWare: Architecting Evolvable Software".

We thank Ian Gent and Tom Kelsey of the St Andrews constraint satisfaction group for helpful discussions on constraint solving, Warwick Harvey for his tutorials on ECLiPSe, and Ron Morrison and Dharini Balasubramaniam for their insight into Architecture Description Languages.

## References

[1] J. O. Kephart and D. M. Chess, "The Vision of Autonomic Computing", IEEE Computer, vol. 36 no. 1, pp. 41-50, 2003.
[2] J. C. Diaz y Carballo, A. Dearle, and R. C. H. Connor, "Thin Servers - An Architecture to Support Arbitrary Placement of Computation in the Internet", Proc. 4th International Conference on Enterprise Information Systems (ICEIS 2002), Ciudad Real, Spain, 2002.
[3] http://www-systems.dcs.st-and.ac.uk/cingal/
[4] D. Garlan, R. Allen, and J. Ockerbloom, "Exploiting Style in Architectural Design Environments", Proc. 2nd ACM SIGSOFT Symposium on Foundations of Software Engineering, New Orleans, Louisiana, USA, 1994.
[5] M. Moriconi, X. Qian, and R. A. Riemenschneider, "Correct Architecture Refinement", IEEE Transactions on Software Engineering, vol. 21 no. 4, pp. 356-372, 1995.
[6] M. Shaw, R. DeLine, D. V. Klein, T. L. Ross, D. M. Young, and G. Zelesnik, "Abstractions for Software Architecture and Tools to Support Them", IEEE Transactions on Software Engineering, vol. 21 no. 4, pp. 314-335, 1995.
[7] D. Garlan, R. Monroe, and D. Wile, "ACME: An Architecture Description Interchange Language", Proc. Conference of the Centre for Advanced Studies on Collaborative Research (CASCON'97), Toronto, Canada, 1997.
[8] R. Morrison, G. N. C. Kirby, D. Balasubramaniam, K. Mickan, F. Oquendo, S. Cîmpan, B. C. Warboys, B. Snowdon, and R. M. Greenwood, "Constructing Active Architectures in the ArchWare ADL", University of St Andrews 2003. http://www.dcs.st-and.ac.uk/research/publications/MKB+03.shtml
[9] R. Keller, J. Ramamirtham, T. Wolf, and B. Plattner, "Active Pipes: Service Composition for Programmable Networks", Proc. IEEE MILCOM 2001, McLean, VA, USA, 2001.
[10] "The ECLiPSe Constraint Logic Programming System", 2003 http://www-icparc.doc.ic.ac.uk/eclipse/
[11] ILOG, "ILOG JSolver", 2004 http://www.ilog.com/products/jsolver/
[12] N. Tamura, "Cream: Class Library for Constraint Programming in Java", 2003 http://bach.istc.kobe-u.ac.jp/cream/
[13] R. Strom and S. Yemini, "Optimistic Recovery in Distributed Systems", ACM Transactions on Computer Systems, vol. 3 no. 3, pp. 204-226, 1985.
[14] A. Dearle, G. N. C. Kirby, A. McCarthy, and J. C. Diaz y Carballo, "A Flexible and Secure Deployment Framework for Distributed Applications", Submitted To 2nd International Working Conference on Component Deployment (CD 2004), 2004.
[15] K. P. Birman and R. Cooper, "The ISIS Project: Real Experience with a Fault Tolerant Programming System", Operating Systems Review, vol. 25 no. 2, pp. 103-107, 1991.
[16] T. Chandra and S. Toueg, "Unreliable Failure Detectors for Reliable Distributed Systems", Journal of the ACM, vol. 43 no. 1, pp. 225-267, 1996.
[17] M. K. Aguilera, W. Chen, and S. Toueg, "Heartbeat: a Timeout-Free Failure Detector for Quiescent Reliable Communication", in Lecture Notes in Computer Science 1320, M. Mavronicolas and P. Tsigas, Eds.: Springer-Verlag, 1997, pp. 126-140.